\documentclass[letterpaper]{article} 
\usepackage{aaai2026}  
\usepackage{times}  
\usepackage{helvet}  
\usepackage{courier}  
\usepackage[hyphens]{url}  
\usepackage{graphicx} 
\usepackage{natbib}  
\usepackage{caption} 
\usepackage{algorithm}
\usepackage{algorithmic}

\usepackage{multirow}
\usepackage{tabularx}
\usepackage{amsmath}
\usepackage{amsfonts}
\usepackage{amssymb}

\usepackage{newfloat}
\usepackage{listings}
\DeclareCaptionStyle{ruled}{labelfont=normalfont,labelsep=colon,strut=off} 
\floatstyle{ruled}
\newfloat{listing}{tb}{lst}{}
\floatname{listing}{Listing}
\title{BeDKD: Backdoor Defense Based on Directional Mapping Module and Adversarial Knowledge Distillation}
\author{
    Zhengxian Wu\textsuperscript{\rm 1},
    Juan Wen\textsuperscript{\rm 1}\thanks{Corresponding authors: Juan Wen and Wanli Peng.},
    Wanli Peng\textsuperscript{\rm 1}\footnotemark[1],
    Yinghan Zhou\textsuperscript{\rm 1},
    Changtong Dou\textsuperscript{\rm 1},
    Yiming Xue\textsuperscript{\rm 1}
}
\affiliations{
    \textsuperscript{\rm 1}College of Information and Electrical Engineering, China Agricultural University, Beijing 100083, China\\
    \{wzxian, wenjuan, wlpeng, zhouyh, ctdou\_sam, xueym\}@cau.edu.cn
}

\usepackage{bibentry}

\begin{document}

\maketitle

\begin{abstract}
Although existing backdoor defenses have gained success in mitigating backdoor attacks, they still face substantial challenges. In particular, most of them rely on large amounts of clean data to weaken the backdoor mapping but generally struggle with residual trigger effects, resulting in persistently high attack success rates (ASR). Therefore, in this paper, we propose a novel \textbf{B}ackdoor d\textbf{e}fense method based on \textbf{D}irectional mapping module and adversarial \textbf{K}nowledge \textbf{D}istillation (BeDKD), which balances the trade-off between defense effectiveness and model performance using a small amount of clean and poisoned data. We first introduce a directional mapping module to identify poisoned data, which destroys clean mapping while keeping backdoor mapping on a small set of flipped clean data. Then, the adversarial knowledge distillation is designed to reinforce clean mapping and suppress backdoor mapping through a cycle iteration mechanism between trust and punish distillations using clean and identified poisoned data. We conduct experiments to mitigate mainstream attacks on three datasets, and experimental results demonstrate that BeDKD surpasses the state-of-the-art defenses and reduces the ASR by 98$\%$ without significantly reducing the CACC.
\end{abstract}

\begin{links}
    \link{Code}{https://github.com/CAU-ISS-Lab/Backdoor-Attack-Defense-LLMs/tree/main/BeDKD}
\end{links}

\section{Introduction}

In recent years, deep neural networks (DNNs) have achieved great success in the field of natural language processing (NLP), such as sentiment analysis \cite{SA2,SA1}, machine translation \cite{MT2,MachineTranslation} and natural language generation \cite{NLG,NLG2}. However, recent studies show that DNNs are highly vulnerable to backdoor attacks \cite{Survey3,Survey,Survey6,Survey7}.

Backdoor attacks generally introduce an invisible vulnerability in DNNs, allowing attackers to control or manipulate the model's output when the input contains the specific trigger patterns \cite{Survey5,Survey4}. To carry out a backdoor attack, the attacker first injects triggers into a small amount of clean data to poison the training set, and then trains the victim model. In inference, the poisoned model responds normally to clean data, while it responds incorrectly to poisoned data based on the attacker's target label. The prevalence of backdoor attacks poses significant security risks to deep neural networks \cite{HD2,HD1,HSP1,HSP2}.

\begin{figure}[t]
\centerline{\includegraphics[width=\linewidth]{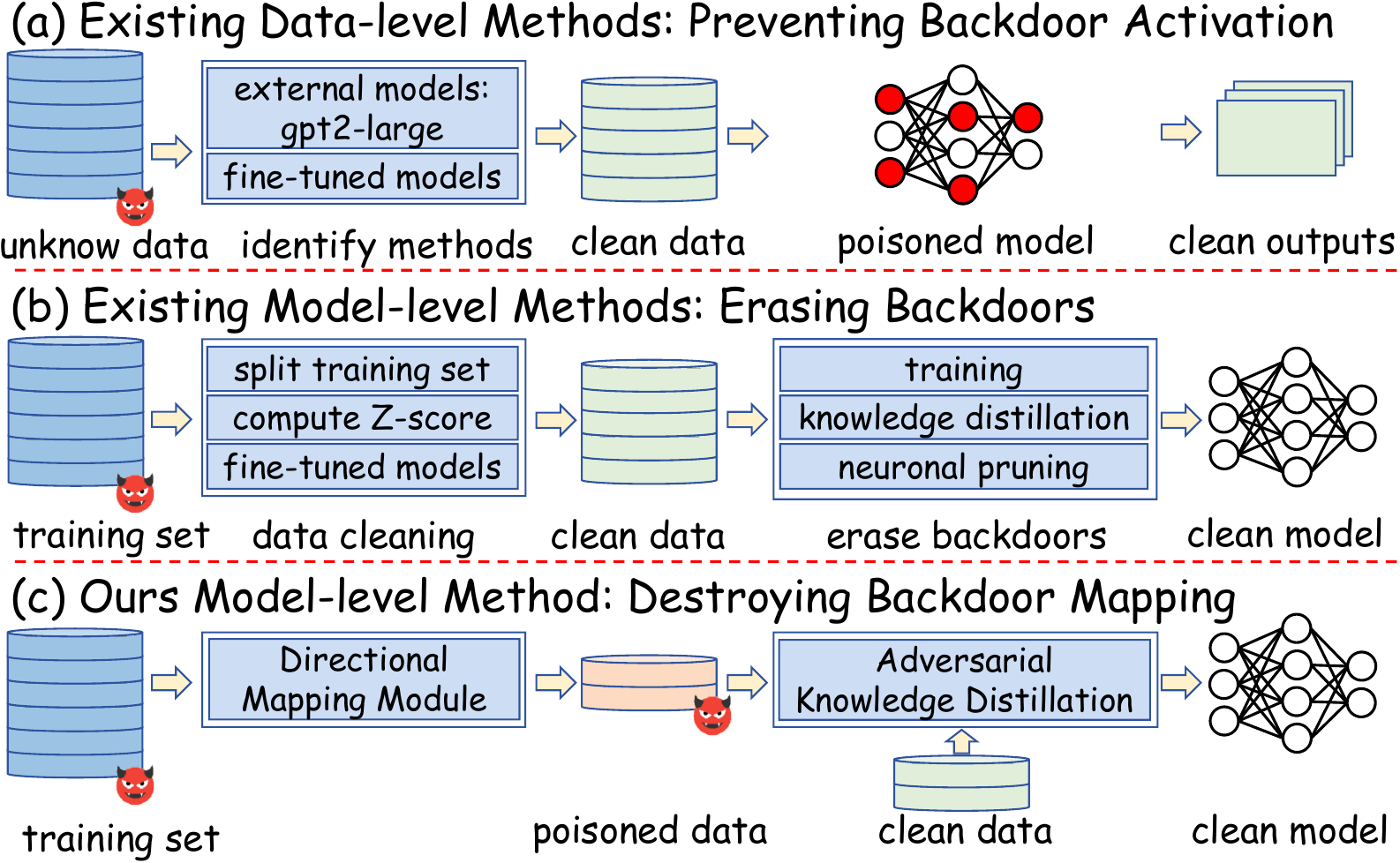}}
\caption{(a) Existing data-level defenses. (b) Existing model-level defenses require sufficient clean data. (c) Our proposed method requires minimal clean and poisoned data.}
\label{fig1}
\end{figure}

To defend against backdoor attacks, researchers have explored many backdoor defense methods, broadly categorized into \textbf{data-level} \cite{BKI,STRIP,mdp-nips23,DefendingInsertion} and \textbf{model-level} \cite{WeDef,PURE,TextGuard} approaches. As shown in Figure \ref{fig1}(a) and (b), the goal of data-level methods is to identify poisoned data, while the goal of model-level methods is to erase the backdoor of the poisoned model. The former identifies poisoned data from the input data via external models or fine-tuned models. Even though these methods have achieved success in mitigating backdoor attacks, their primary strategy is to avoid activating backdoors rather than essentially eliminate backdoors. In contrast, the later mainly erases backdoors through data cleaning, training, knowledge distillation (KD), or neuronal pruning. Although the existing model-level methods remove backdoors effectively, they reduce the accuracy of the poisoned model on the clean data. \textbf{\textit{Therefore, achieving a satisfactory trade-off between backdoor defense and maintaining model performance remains a significant challenge.}}

More recently, some defense methods have been introduced to alleviate the above trade-off problem. Zhao et al. \cite{PEFT} randomly flips the label of a clean proxy dataset to fine-tune the poisoned model, enabling it to identify poisoned data. To erase backdoors, Zhao et al. \cite{PEFT-Distillation} leverages a clean proxy dataset to fine-tune the BERT and uses the fine-tuned BERT as the teacher model, which guides the poisoned student model to unlearn the backdoors via knowledge distillation. Although they excel at both mitigating backdoor attacks and preserving model performance, they require quantities of clean data to fine-tune models, limiting their application in the real world.

From the above analysis, in this paper, we explore a novel model-level \textbf{B}ackdoor d\textbf{e}fense method based on a \textbf{D}irectional mapping module and adversarial \textbf{K}nowledge \textbf{D}istillation (BeDKD). Typically, the poisoned model has two mappings: clean mapping and backdoor mapping. Clean mapping is the correlation between the semantics of clean data and ground-truth labels, while backdoor mapping refers to the relationship between triggers and the target label. Intuitively, backdoor erasing is equivalent to destroying the backdoor mapping while maintaining the clean mapping. Different from existing backdoor defenses that utilize clean data to weaken the backdoor mapping, we employ poisoned data to break the backdoor mapping. Specifically, BeDKD (as shown in Figure \ref{fig1}(c)) employs a directional mapping module to effectively identify poisoned data and then utilizes the adversarial knowledge distillation to preserve clean mapping while enforcing suppression of backdoor mappings using small subsets of clean and poisoned data.

Most of existing defenses rely on large amounts of clean data, making it difficult to adapt to real-world scenarios with limited clean data. Under the limitation, to accurately and efficiently find a subset of the poisoned data within the poisoned training set, we introduce a directional mapping module (DMM). The DMM, which copies the architecture and parameters of the poisoned model, is fine-tuned on a small number of clean data with intentionally flipped labels to disrupt the clean mapping. By analyzing the distribution's difference between the poisoned model and the fine-tuned DMM, the poisoned data can be effectively identified.

Due to the robust retention of trigger features and the concealment of backdoor trigger design, existing methods only using clean data to defend against backdoor attacks generally suffer from trigger residue, resulting in high attack success rate (ASR). Therefore, we propose a adversarial knowledge distillation (AKD), which employs a cycle iteration mechanism to maintain the clean mapping and erase the backdoor mapping using a small amount of clean and poisoned data. Each AKD cycle iteration consists of two stages: trust distillation and punish distillation. The former leverages a small set of clean data to enable the student model to learn clean mapping from the teacher model, while the latter enables the student model to erase backdoor mapping on a handful of poisoned data through a penalty loss function.

We conduct extensive experiments on SST2, OLID, and AGnews to evaluate the performance of our proposed BeDKD. Extensive experimental results demonstrate that our proposed method can reduce ASR by 98$\%$ and without significantly compromising CACC in most cases, which outperforms the state-of-the-art backdoor defense methods.

Our contributions are summarized as follows:
\begin{itemize}
\item We explore a novel model-level backdoor defense based on directional mapping module and adversarial knowledge distillation (BeDKD), which makes a satisfied trade-off between defense effectiveness and model performance via a small amount of clean and poisoned data.
\item We introduce a directional mapping module (DMM) that destroys clean mapping from a handful of clean data through transfer learning to identify poisoned data. To suppress backdoor mapping, the adversarial knowledge distillation (AKD) is designed, which guides the poisoned student model to learn clean mapping on clean data through trust distillation and push away backdoor mapping on poisoned data through punish distillation from the poisoned teacher model.
\item We conduct extensive experiments to evaluate the effectiveness of our method on three public benchmarks: OLID, SST2, and AGnews. Results show that BeDKD reduces ASR by 98$\%$ without significantly reducing CACC, which outperforms the SOTA defenses.
\end{itemize}

\section{Related Work}
\subsection{Backdoor Attack}\label{2.1}
Dai et al. \cite{AddSent} and Chen et al. \cite{BadWord} insert meaningful fixed short sentences and rare words into clean data. To improve the stealthiness of triggers, Qi et al. \cite{SynBkd} and Pan et al. \cite{StyBkd} rewrite sentences with a specific syntactic structure and style. Yan et al. \cite{BITE} capitalize on spurious correlations between the target label and specific words in training data. To further improve stealthiness and text quality, Du et al. \cite{AIGTBackdoor} fine-tune LLMs based on attribute control to generate poisoned data. Similarly, Li et al. \cite{BGMAttack} design hand-crafted prompt to guide LLMs to generate rephrased poisoned data. With the advancement of backdoor attacks, designing an accurate and effective backdoor defense is still a critical challenge.

\begin{figure*}[t]
\centerline{\includegraphics[width=0.93\textwidth]{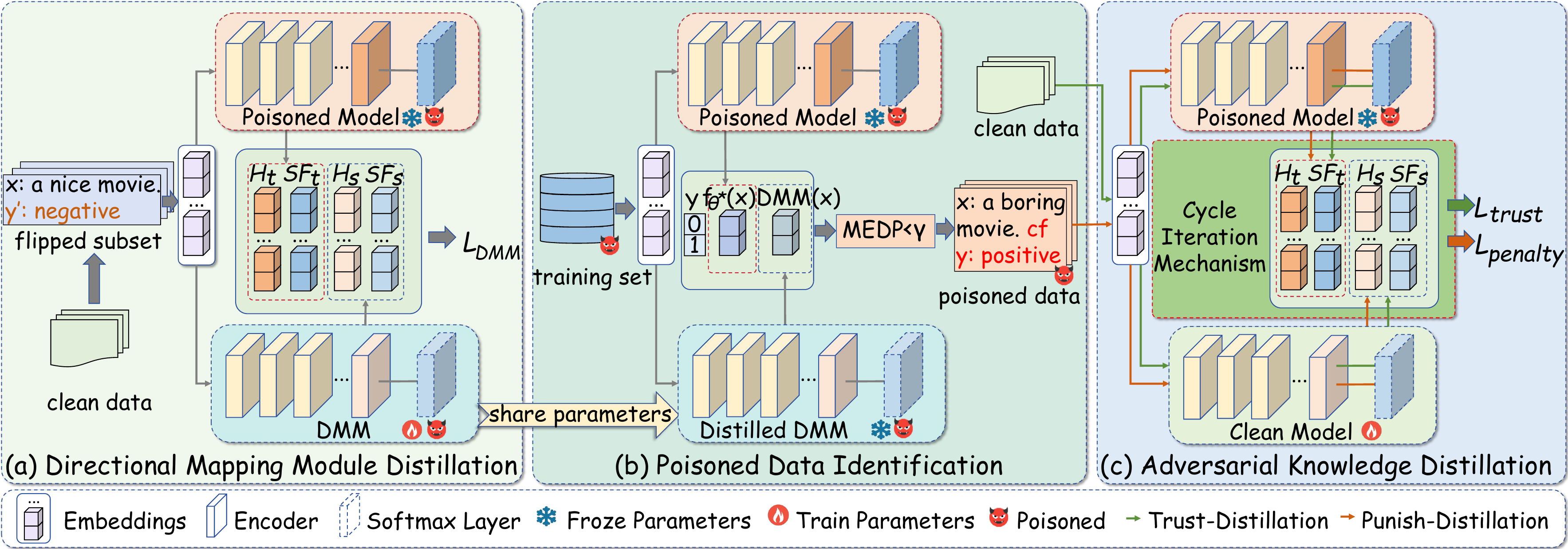}}
\caption{Our BeDKD framework. (a) Directional mapping module distillation. We distill the DMM from the poisoned model ($f_{\theta^{*}}$) on the flipped data, a small number of clean data with flipped labels, to destroy the clean mapping. (b) Poisoned data identification. We compute the mean error of probability distributions (MEPD) between the $f_{\theta^{*}}$ and the distilled DMM to identify a handful of poisoned data from the poisoned training set. (c) Adversarial knowledge distillation. The $f_{\theta^{*}}$ guides the poisoned student model (CM) to pull the clean mapping on the clean data and push away the backdoor mapping on the poisoned data via a cycle iteration mechanism, which alternates trust and punish distillations. Notably, the initial DMM and CM have the same architecture and parameters as $f_{\theta^{*}}$.}
\label{fig2}
\end{figure*}
\subsection{Backdoor Defense} \label{2.2}
(1) $\textbf{Data-Level Defenses.}$ Qi et al. \cite{ONION} utilize an external language model as a grammar outlier detector to remove trigger words from the input. Yang et al. \cite{RAP} use an additional prompt-based optimizer to verify the output logit permutation. Chen et al. \cite{BKI} identify trigger words using word importance scores. Due to the poisoned model's sensitivity to triggers, Gao et al. \cite{STRIP} detect poisoned data by randomly perturbing features and analyzing output changes of each data. Similarly, He et al. \cite{IMBERT} used gradients or self-attention scores to self-defend against backdoor attacks. Although existing data-level defenses successfully defend against backdoor attacks, they still have live backdoors. (2) $\textbf{Model-Level Defenses.}$ He et al. \cite{Zscore} compute the spurious correlation between text features and labels to clean the poisoned training set and retain the victim model. Zhao et al. \cite{PURE} erase backdoors through attention head pruning and weight- normalization. Pei et al. \cite{TextGuard} train multiple classifiers on divided $m$ sub-training sets and ensemble their predictions. These defenses mitigate backdoor attacks effectively, while they struggle to balance the defense trade-off and require substantial clean data for fine-tuning.
\subsection{Knowledge Distillation}\label{2.3}
Knowledge distillation (KD) compresses larger or ensemble networks (teacher models) into smaller networks (student models) \cite{KD1}. Feature maps and attention mechanisms have proven effective in KD, enabling student models to learn high-quality intermediate representations from teacher models, thereby enhancing distillation and improving performance \cite{KD2,KD3}. KD has been applied to speech recognition \cite{SRKD1,SRKD2}, visual recognition \cite{KD4,VRKD2}, backdoor defense \cite{NAD,PEFT-Distillation}. Zhao et al. \cite{PEFT-Distillation} fine-tune BERT on a large task-related clean dataset as the teacher model to guide the poisoned model to erase backdoors via knowledge distillation. However, they rely heavily on large volumes of clean data, posing challenges in low-resource scenarios.
\section{Methodology}
\subsection{Preliminaries} \label{Preliminaries}
\subsubsection{Attacker's Goal.} Attackers contaminate the training sets and upload them to third-party platforms (e.g., HuggingFace, GitHub, etc.). When users train or fine-tune models on these sets, the backdoor mapping is automatically introduced into the victim models. Specifically, attackers divide the training set $D$ into two subsets: $D_c$, which is reserved as clean data, and $D_p$, which is used for poisoning. Then, a transform operation $F:\{(x,y)\rightarrow(x^{*},y_t)\}$ is designed, where $x$ is the clean sample, $y$ is the corresponding label, $x^{*}$ represents the poisoned sample obtained by inserting trigger $t$ into the clean sample $x$, and $y_{t}$ represents the target label. The operation $F$ is applied to $D_{p}$ to obtain the poisoned subset $D_{p}^{*}$. The optimization objectives of the victim model 
are $\mathrm {\theta^{*}} = \underset{\theta }{arg\ min} \{\mathbb{E}_{(x_{i},y_{i})\sim D_{c}}[\mathcal{L}(f_{\theta}(x_{i}),y_{i}) ]\ +\ \mathbb{E}_{(x_{i}^{*} ,y_{t})\sim D_{p}^{*}}[\ \mathcal{L}(f_{\theta}(x_{i}^{*}),\ y_{t}) ]\}$, where $\theta$ is the parameter of the victim model $f$. $\mathcal{L}$ is the cross-entropy loss function. The poisoned model only activates backdoor mapping on triggered inputs and maintains normal mapping on clean inputs.
\subsubsection{Defender's Goal.} Following the previous backdoor defenses \cite{BKI, TextGuard,PEFT-Distillation}, the defender is user. The defender has access to the training set but is unaware of the presence of poisoned data within it. The goal of defender is to distill a clean model using the downloaded poisoned dataset, while preserving the clean mapping and eliminating the backdoor mapping. This means that the defended model should have a low attack success rate on the poisoned test set, while maintaining a high classification accuracy on the clean test set.

\subsection{Overview of BeDKD}
Figure \ref{fig2} illustrates the framework of our proposed BeDKD, which consists of three key steps: directional mapping module (DMM) distillation, poisoned data identification, and adversarial knowledge distillation (AKD). First, the DMM is distilled on a small flipped clean samples to enhance the backdoor mapping, after which it identifies a small amount of poisoned data from the training set. Then, the AKD is applied to derive a clean model from the poisoned model, using both the identified poisoned data and a small amount of clean data, following a cycle iteration mechanism.


\subsection{Distilled DMM for Locating Poisoned Data}\label{DMM}

Traditional backdoor defenses use clean data for fine-tuning or distillation to erase the backdoors \cite{PURE,PEFT-Distillation}. However, they require a large number of clean data and fall short of completely eliminating the backdoor mapping (higher ASR). This paper leverages a small number of clean samples to identify a small number of poisoned samples and incorporates them into the distillation process, enabling the model to more effectively remove backdoors. To find poisoned samples, we propose the Directional Mapping Module (DMM), which has the same structure as the poisoned model and is distilled by a small amount of flipped clean data to disrupt the clean mapping of the DMM while reinforcing the backdoor mapping, thereby facilitating the identification of trustworthy poisoned samples. The goal of DMM is to make the probability distribution difference of clean mapping as large as possible, while making the probability distribution difference of backdoor mapping as small as possible.


Assume that we have access to a small number of clean data $D_{c}^{few}$ \cite{Fine-tuning,PURE,PEFT-Distillation}. We modify the ground-truth label $y$ of clean data $x$ and flip it to an incorrect label $y'\in Y$ to create a flipped clean data $D_{c}^{few'}$, where $Y$ is label space. We initialized the DMM with shared parameters from the $f_{\theta^{*}}$. 

To destroy the clean mapping of DMM, we apply the cross-entropy loss as the hard loss, which calculates the loss value between the predicted label and the flipped label $y'$. The formula is as follows:
\begin{equation}
    \small L_{hard}=- {\textstyle \sum_{(x,y')\in D_{c}^{few'}} y'\log(DMM(x))},
    \label{hard}
\end{equation}
where, $DMM(\cdot)$ is the prediction of the DMM.

Fine-tuning the DMM on the flipped data $D_{c}^{few'}$ is equivalent to introducing a new mapping relationship, which leads the DMM to readjust the feature distribution and reduces the stability of backdoor mapping. To reinforce the backdoor mapping of DMM, we introduce knowledge distillation for feature alignment by incorporating Kullback-Leibler (KL) divergence and mean square error (MSE) loss as soft loss:
\begin{equation}
\begin{aligned}
     L_{soft}= - {\textstyle \sum_{x\in D_{c}^{few'}} SF_t(x,T)\log(SF_s(x,T))} \\
     + Mean(\textstyle \sum_{x\in D_{c}^{few'}} (H_t(x)-H_s(x)^{2}),
\end{aligned}
     \label{soft}
\end{equation}
where $T$ is the temperature. $SF_t(x,T)$ and $SF_s(x,T)$ are the softmax layer output of the poisoned teacher $f_{\theta^{*}}$ and student model DMM with $T$, respectively. $H_t(\cdot)$ and $H_s(\cdot)$ are the last hidden sates of $f_{\theta^{*}}$ and DMM, respectively.

In the fine-tune stage of DMM, the total loss is formulated by combining the hard loss (Eq.\ref{hard}) and soft loss (Eq.\ref{soft}) to achieve the desired balance between disrupting the clean mapping and preserving the backdoor mapping. The total loss is as follows:
\begin{equation}
    L_{DMM}= \alpha L_{hard} +(1-\alpha)*(L_{soft}),
    \label{LDMM}
\end{equation}
where $\alpha \in [0,1]$ is the hyper-parameter.

After distilling the DMM, there will be a deviation in the probability distribution for clean inputs between the DMM and $f_{\theta^{*}}$, while the output probabilities for poisoned inputs show almost no deviation. Therefore, poisoned data can be identified by calculating the mean error of the probability distributions between the DMM and $f_{\theta^{*}}$.
\begin{equation}
    \small MEPD = \frac{{\textstyle \sum_{y}^{Y}} abs(f_{\theta^{*}}(x,y)- DMM(x,y))}{|Y|} ,
    \label{MEDP}
\end{equation}
where $abs(\cdot)$ is the absolute value function. $f_{\theta^{*}}(x,y)$ represents the probability that the data $x$ is predicted to be $y$. When the MEPD of the data is less than the threshold $\gamma$, it is considered to be poisoned. Otherwise, it is clean data. The $\gamma$ is determined through a small number of clean data.
\subsection{Adversarial Knowledge Distillation}\label{AKD}

Traditional knowledge distillation focuses on guiding the student model to learn the feature distributions of the teacher model, thereby facilitating knowledge transfer and enhancing generalization \cite{Distillation}. However, in backdoor defense task, directly applying traditional knowledge distillation can lead the student model to simultaneously learn both the clean and backdoor mapping from the poisoned teacher model, making it difficult to eliminate backdoors (detailed discussion in Section \ref{Ablations}). In addition, although some studies utilize task-related clean datasets to distill a clean model from the poisoned model, such as W2SDefense \cite{PEFT-Distillation}, they require a large amount of clean data, which limits their practical application. To address this issue, we propose an Adversarial Knowledge Distillation (AKD), which employs an adversarial distillation strategy to promote the learning of clean mapping while suppressing backdoor mapping on limited clean and poisoned data (as shown in Figure \ref{fig2}(c)). Specifically, the teacher model is the poisoned model $f_{\theta^{*}}$ with frozen parameters, while the student model ($CM$) shares the same architecture and parameters as $f_{\theta^{*}}$. The AKD adopts a cycle iteration mechanism, performing trust distillation on a small amount of clean data and punish distillation on a small amount of poisoned data identified in the previous step. By alternating between two types of distillation, the backdoor mapping is eliminated without reducing the clean mapping.


To be specific, trust distillation utilizes the clean data $D_{c}^{few}$ to instruct the $CM$ reinforce the learning of clean mapping from the $f_{\theta^{*}}$. The loss function is shown below:
\begin{equation}
    L_{trsut}= \lambda L_{hard} +(1-\lambda)*(L_{soft}),
    \label{LAKD}
\end{equation}
where $\lambda$ is the hyper-parameter.

Punish distillation applies a small number of poisoned data $D_{p}^{few*}$ identified by DMM to prevent the $CM$ from learning the backdoor mapping of the $f_{\theta^{*}}$ to erase the backdoor via the penalty loss function. The loss function:
\begin{equation}
    L_{penalty}= -(\lambda L_{hard} +(1-\lambda)*(L_{soft})).
    \label{-LAKD}
\end{equation}

\begin{table*}[t]
\small
\centering
\setlength{\tabcolsep}{1mm}       
\begin{tabular}{ccccccccccccccc}
\hline
\multirow{2}{*}{Attacks} & \multicolumn{2}{c}{No Defense} & \multicolumn{2}{c}{FT} & \multicolumn{2}{c}{ONION} & \multicolumn{2}{c}{IMBERT} & \multicolumn{2}{c}{TextGuard} & \multicolumn{2}{c}{W2SDefense} & \multicolumn{2}{c}{Ours}       \\ \cline{2-15}
                         & ASR$\uparrow$           & CACC$\uparrow$         & ASR$\downarrow$        & CACC$\uparrow$             & ASR$\downarrow$        & CACC$\uparrow$             & ASR$\downarrow$            & CACC$\uparrow$          & ASR$\downarrow$         & CACC$\uparrow$                & ASR$\downarrow$          & CACC$\uparrow$          & ASR$\downarrow$          & CACC$\uparrow$          \\ \hline
\multicolumn{15}{c}{SST2}                                                                                                                                                                                                                                       \\ \hline
Clean                    & -              & 91.97         & -           & 89.79             & -           & \underline{90.02}       & -               & 83.95          & -            & 89.45                & -             & 89.91          & -             & \textbf{91.06} \\
BadWords                 & 100.00         & 91.63         & 63.06       & 88.65             & 49.32       & 89.40             & \underline{20.95}     & 83.95          & 35.59        & 89.56                & 21.17         & \underline{89.79}    & \textbf{0.00} & \textbf{90.14} \\
AddSent                  & 100.00         & 91.62         & 72.07       & 88.07             & 91.67       & 88.07             & \underline{18.02}     & 85.67          & 21.40        & \underline{90.02}          & 55.63         & \textbf{91.17} & \textbf{0.00} & \textbf{91.17} \\
Syntax                   & 95.27          & 91.51         & 66.22       & 89.22             & 90.09       & 90.02             & 89.86           & 86.01          & 48.42        & 89.11                & \underline{40.09}   & \textbf{90.71} & \textbf{2.48} & \underline{90.48}    \\
StyBkd                   & 85.14          & 90.14         & 55.50       & 89.79             & 68.92       & 85.21             & 82.88           & 81.77          & 70.72        & 82.34                & \underline{27.48}   & \underline{90.25}    & \textbf{4.86} & \textbf{90.59} \\
AttrBkd                  & 95.95          & 91.86         & 95.05       & \underline{90.48}       & 95.50       & 88.19             & 96.17           & 89.11          & 96.62        & 87.04                & \underline{4.96}    & \textbf{91.28} & \textbf{0.23} & \underline{90.48}    \\
BGMAttack                & 99.32          & 83.14         & 47.30       & \underline{86.47}       & 93.07       & 67.91             & 95.16           & 73.37          & 88.71        & 77.79                & \underline{14.19}   & \textbf{90.25} & \textbf{3.15} & \textbf{90.25} \\
Average                  & 95.95          & 90.27         & 66.53       & 88.92             & 81.43       & 85.55             & 67.17           & 83.40          & 60.24        & 86.47                & \underline{27.25}   & \underline{90.48}    & \textbf{1.79} & \textbf{90.60} \\ \hline
\multicolumn{15}{c}{OLID}                                                                                                                                                                                                                                                \\ \hline
Clean                    & -              & 82.79         & -           & \underline{83.14}       & -           & 81.98             & -               & 80.58          & -            & \textbf{84.19}       & -             & 80.70          & -             & 81.39          \\
BadWords                 & 100.00         & 83.95         & 92.08       & 79.30             & 79.17       & 80.93             & 82.08           & \underline{82.33}    & 59.58        & \textbf{84.07}       & \underline{10.83}   & 79.30          & \textbf{0.00} & 80.81          \\
AddSent                  & 100.00         & 81.98         & 95.83       & 79.88             & 95.00       & \underline{82.09}       & 85.42           & 81.51          & 100.00       & \textbf{84.88}       & \underline{6.25}    & 79.42          & \textbf{0.00} & 81.28          \\
Syntax                   & 99.58          & 82.67         & 96.25       & 81.28             & 98.75       & 80.35             & 98.33           & \underline{82.33}    & 96.67        & \textbf{83.95}       & \underline{10.00}   & 80.70          & \textbf{1.67} & 79.88          \\
StyBkd                   & 92.58          & 79.65         & 76.61       & 80.00             & 91.61       & 73.26             & 96.45           & 82.91          & 87.42        & \underline{83.26}          & \underline{47.10}   & 80.35          & \textbf{2.90} & \textbf{84.30} \\
AttrBkd                  & 97.42          & 78.95         & 82.91       & 75.35             & 80.48       & 77.44             & 96.77           & 75.47          & 97.74        & 77.58                & \underline{10.65}   & \underline{78.37}    & \textbf{2.26} & \textbf{82.79} \\
BGMAttack                & 97.26          & 73.95         & 62.75       & 78.26             & 93.07       & 67.91             & 95.16           & 73.37          & 88.71        & 77.79                & \underline{20.81}   & \underline{79.65}    & \textbf{0.81} & \textbf{83.37} \\
Average                  & 97.81          & 80.56         & 84.41       & 79.60             & 89.68       & 77.71             & 92.37           & 79.79          & 88.35        & \textbf{82.25}       & \underline{17.61}   & 79.78          & \textbf{1.27} & \underline{81.97}    \\ \hline
\multicolumn{15}{c}{AGnews}                                                                                                                                                                                                                                              \\ \hline
Clean                    & -              & 93.96         & -           & 92.87             & -           & 92.33             & -               & \underline{93.12}    & -            & 91.93                & -             & \textbf{93.93} & -             & 92.86          \\
BadWords                 & 100.00         & 94.01         & 51.09       & 92.47             & 29.65       & 91.97             & 12.30           & 93.13          & 63.32        & 91.65                & \underline{1.67}    & \textbf{93.94} & \textbf{0.04} & \underline{93.53}    \\
AddSent                  & 100.00         & 93.90         & 43.46       & 92.43             & 65.75       & 91.86             & 11.81           & 93.01          & 2.18         & 91.65                & \textbf{0.00} & \textbf{93.92} & \textbf{0.00} & \underline{93.53}    \\
Syntax                   & 99.88          & 93.92         & 35.16       & 92.83             & 94.91       & 91.18             & 94.37           & 92.55          & 5.75         & 91.75                & \underline{0.39}    & \underline{93.91}    & \textbf{0.02} & \textbf{94.00} \\
StyBkd                   & 97.33          & 92.90         & 71.05       & 92.78             & 98.19       & 90.38             & 96.86           & 92.24          & 56.82        & 85.58                & \underline{10.37}   & \textbf{94.07} & \textbf{2.28} & \underline{93.51}    \\
AttrBkd                  & 98.70          & 93.30         & 91.05       & 92.17             & 97.68       & 91.50             & 98.32           & 92.45          & 97.87        & 88.34                & \underline{2.42}    & \textbf{93.82} & \textbf{0.42} & \underline{93.68}    \\
BGMAttack                & 99.25          & 93.40         & 70.98       & 92.49             & 70.49       & 91.16             & 93.63           & 92.76          & 98.92        & 69.97                & \underline{5.21}    & \textbf{94.05} & \textbf{2.12} & \underline{93.50}    \\
Average                  & 99.19          & 93.63         & 60.47       & 92.58             & 76.11       & 91.48             & 67.88           & 92.75          & 54.14        & 87.27                & \underline{3.34}    & \textbf{93.95} & \textbf{0.81} & \underline{93.52}   \\
\hline
\end{tabular}

\caption{ASR and CACC of the proposed method compare with baselines. The $\textbf{bold}$ and \underline{underline} are the best and second best values. "Clean" means the performance of clean model, which trains on clean dataset. }
\label{baselines}
\end{table*}

\begin{table*}[t]
\centering
\small
\setlength{\tabcolsep}{1.8mm}       
\begin{tabular}{ccccccccccccc}
\hline
\multirow{2}{*}{Defenses} & \multicolumn{2}{c}{BadWords} & \multicolumn{2}{c}{AddSent} & \multicolumn{2}{c}{SynBkd} & \multicolumn{2}{c}{StyBkd} & \multicolumn{2}{c}{AttrBkd} & \multicolumn{2}{c}{BGMAttack} \\ \cline{2-13}
                          & ASR$\downarrow$          & CACC$\uparrow$         & ASR$\downarrow$          & CACC$\uparrow$        & ASR$\downarrow$         & CACC$\uparrow$        & ASR$\downarrow$         & CACC$\uparrow$        & ASR$\downarrow$         & CACC$\uparrow$         & ASR$\downarrow$          & CACC$\uparrow$          \\ \hline
FT                        & 63.06         & 88.65        & 72.07         & 88.07       & 66.22        & 89.22       & 55.50        & 89.79       & 95.05        & 90.48        & 47.30         & 86.47         \\
FT+DMM                    & 19.60         & 88.30        & 10.59         & 89.33       & 22.97        & 87.84       & 37.39        & 88.65       & 25.90        & 89.83        & 29.73         & 89.79         \\ \hline
KD                        & 100.00        & 91.74        & 100.00        & 91.40       & 94.60        & 91.97       & 69.60        & 90.48       & 95.72        & 91.28        & 97.52         & 86.58         \\
KD+DMM                    & 20.50         & 91.86        & 14.41         & 91.63       & 40.54        & 91.17       & 49.78        & 90.60       & 65.77        & 89.91        & 68.69         & 90.14         \\
AKD+DMM                   & 0.00          & 90.14        & 0.00          & 91.17       & 2.48         & 90.48       & 4.86         & 90.59       & 0.23         & 90.48        & 3.15          & 90.25       \\
\hline
\end{tabular}

\caption{Performance of DMM and ADK on the SST2.}
\label{ablation}
\end{table*}

The optimize objectives of AKD as follows: 
\begin{multline}
 \tilde{\theta^{*}}=\underset{\theta^* }{arg\ min}\{\mathbb{E}_{(x_{i},y_{i})\sim D_{c}^{few}}[\mathcal{L}_{trust}(f_{\theta^*}(x_{i}),y_{i}) ] \\
 +\mathbb{E}_{(x_{i},y^*)\sim D_{p}^{few*}}[\mathcal{L}_{penalty}(f_{\theta^*}(x_{i}),y^*) ]\}. 
\end{multline}

During the training stage, the AKD performs a cycle iteration mechanism, alternating between trust and punish distillation. By alternating these two distillations, AKD ensures that the clean mapping is strengthened through the trust distillation, while the backdoor mapping is gradually erased during the punish distillation. 

\section{Evaluation}
\subsection{Evaluation Settings}
\subsubsection{Datasets \& Attacks.}
\ We conduct experiments on SST2 \cite{SST2}, AGnews \cite{AGNEWS}, and OLID \cite{OLID}. We simulate six prominent backdoor attacks: \textbf{AddSent} \cite{AddSent}, \textbf{BadWords} \cite{BadWord}, \textbf{SynBkd} \cite{SynBkd}, \textbf{StyBkd} \cite{StyBkd}, \textbf{AttrBkd} \cite{AIGTBackdoor}, and \textbf{BGMAttack} \cite{BGMAttack}.
\subsubsection{Baselines \& Metrics. }
\ We compare BeDKD with five mainstream defenses: \textbf{Fine-Tuning (FT)} \cite{Fine-tuning}, \textbf{ONION} \cite{ONION}, \textbf{IMBERT} \cite{IMBERT}, \textbf{TextGuard} \cite{TextGuard}, and \textbf{W2SDefense} \cite{PEFT-Distillation}.
To be fair, we follow previous studies and utilize four commonly adopted metrics. \textbf{ASR} and \textbf{CACC} measure the accuracy of poisoned data and clean data. \textbf{FAR} means the percentage of poisoned data classified as clean out of all poisoned data. \textbf{FRR} means the percentage of clean data classified as poisoned out of all clean data.

\begin{table*}[t]
\centering
\small
\setlength{\tabcolsep}{1.8mm}       
\begin{tabular}{ccccccccccccc}
\hline
\multirow{2}{*}{Loss Functions} & \multicolumn{2}{c}{BadWords} & \multicolumn{2}{c}{AddSent} & \multicolumn{2}{c}{SynBkd} & \multicolumn{2}{c}{StyBkd} & \multicolumn{2}{c}{AttrBkd} & \multicolumn{2}{c}{BGMAttack} \\ \cline{2-13}
                      & FAR          & FRR         & FAR          & FRR        & FAR         & FRR        & FAR         & FRR        & FAR         & FRR         & FAR          & FRR          \\ \hline
$L_{hard}$                 & 0.00          & 4.13         & 76.13         & 4.59        & 66.67        & 3.90        & 59.60        & 19.95       & 99.55        & 17.09        & 67.58         & 32.00         \\
$L_{hard}$+$L_{soft}$                  & 0.00          & 0.92         & 0.00          & 0.69        & 42.12        & 1.38        & 50.23        & 1.49        & 6.08         & 2.06         & 29.28         & 1.03         \\ \hline
\end{tabular}
\caption{Performance of the loss function in DMM.}
\label{ablation2}
\end{table*}

\begin{table*}[t]
\centering
\small
\setlength{\tabcolsep}{2mm}       
\begin{tabular}{lllllllllllll}
\hline
\multirow{2}{*}{n$_{c}$} & \multicolumn{2}{l}{BadWords} & \multicolumn{2}{l}{AddSent} & \multicolumn{2}{l}{SynBkd} & \multicolumn{2}{l}{StyBkd} & \multicolumn{2}{l}{AttrBkd} & \multicolumn{2}{l}{BGMAttack} \\ \cline{2-12}
                         & ASR$\downarrow$         & CACC$\uparrow$         & ASR$\downarrow$         & CACC$\uparrow$        & ASR$\downarrow$        & CACC$\uparrow$        & ASR$\downarrow$         & CACC$\uparrow$       & ASR$\downarrow$         & CACC$\uparrow$        & ASR$\downarrow$          & CACC$\uparrow$         \\ \hline
80                       & 0.90         & 90.60         & 0.00         & 88.19        & 0.90        & 85.09        & 1.94         & 87.84       & 0.00         & 85.44        & 1.38          & 88.53         \\
160                      & 0.00         & 91.86         & 0.00         & 90.25        & 2.70        & 89.45        & 2.65         & 88.32       & 0.23         & 90.48        & 2.70          & 89.91         \\
320                      & 0.00         & 90.14         & 0.00         & 91.17        & 2.48        & 90.48        & 4.86         & 90.59       & 0.23         & 90.48        & 3.15          & 90.25         \\
640                      & 0.23         & 89.91         & 0.00         & 91.40        & 4.51        & 89.33        & 10.09        & 90.71       & 0.68         & 91.63        & 4.96          & 91.28        \\
\hline
\end{tabular}

\caption{CACC and ASR of different scale of clean data on the SST2. $n_c$ is the number of clean samples in each class.}
\label{sensitivety}
\end{table*}

\subsubsection*{Implementation Details.}
\ We leverage the AdamW optimizer with the learning rate of 3 $\times$ $10^{-5}$ to train the poisoned model (widely used BERT) for 10 epochs. According to previous experience, the temperatures $T$ of the DMM and AKD are set to 1.5 and 2.5, respectively. The $\alpha$ and $\lambda$ are both set to 0.3. We train the DMM and AKD for 20 epochs and 50 epochs.

\subsection{Comparison Results}
Table \ref{baselines} summarizes the performance comparison of BeDKD with baselines. "No Defense" means the poisoned models without any defenses. All backdoor attacks always achieve more than 99$\%$ ASR. BeDKD significantly outperforms baselines on most attack settings and lowers around 98$\%$ of all backdoor attacks without compromising CACC in most cases. For insertion-based attacks, BadWords and AddSent use visible rare words and fixed sentences as triggers. Although most baselines can mitigate them, BeDKD achieves lower ASR and higher CACC, especially the average ASR and CACC on three datasets achieve 0.01$\%$ and 88.41$\%$, which is better than the best baseline, W2SDefense (average ASR 15.92$\%$ and CACC 87.83$\%$). For paraphrase-based attacks, SynBkd, StyBkd, AttrBkd, and BGMAttack use invisible syntax templates, style, attribution, and AI-generated texts as triggers. BeDKD still reduces the average ASR to 1.93$\%$ ($\downarrow$16.14$\%$ than W2SDefense). These results show that BeDKD effectively mitigates both visible and invisible triggers. On the OLID dataset, all defense baselines cannot work well because the small scale of the dataset. While BeDKD still effectively defends against all backdoor attacks on OLID dataset and reduces the average ASR to 1.27$\%$ ($\downarrow$17.61$\%$ than W2SDefense). Moreover, BeDKD requires 232s to achieve defense on SST2, which is dramatically more efficient, running orders of magnitude faster than W2SDefense ($>$4,000s) and TextGuard ($>$30,000s). Overall, BeDKD makes a satisfactory trade-off on a small amount of clean data. 

\subsection{Ablation Study}\label{Ablations}
\subsubsection*{The Impact of DMM and AKD. } 
\ Table \ref{ablation} shows that both the DMM and AKD significantly enhance the effectiveness of defense. The FT and KD methods both suffer from trigger residue, where they only reduce the average ASR to almost 66.53$\%$ and 92.91$\%$, respectively. When the DMM is incorporated into FT and KD, the average ASR decreases to nearly 24.36$\%$ and 43.28$\%$, while the CACC remains unchanged. Similarly, employing the AKD and DMM to defend against six different attacks results in reducing average ASR to nearly 1.79$\%$, with CACC only decreasing almost 1$\%$. This indicates that the AKD effectively erases the backdoor mapping to the maximum extent while preserving the clean mapping. Consequently, our proposed BeDKD, which integrates the DMM and AKD, achieves the lowest ASR while maintaining acceptable CACC.

\begin{figure}[t]
	\centerline{\includegraphics[width=0.93\linewidth]{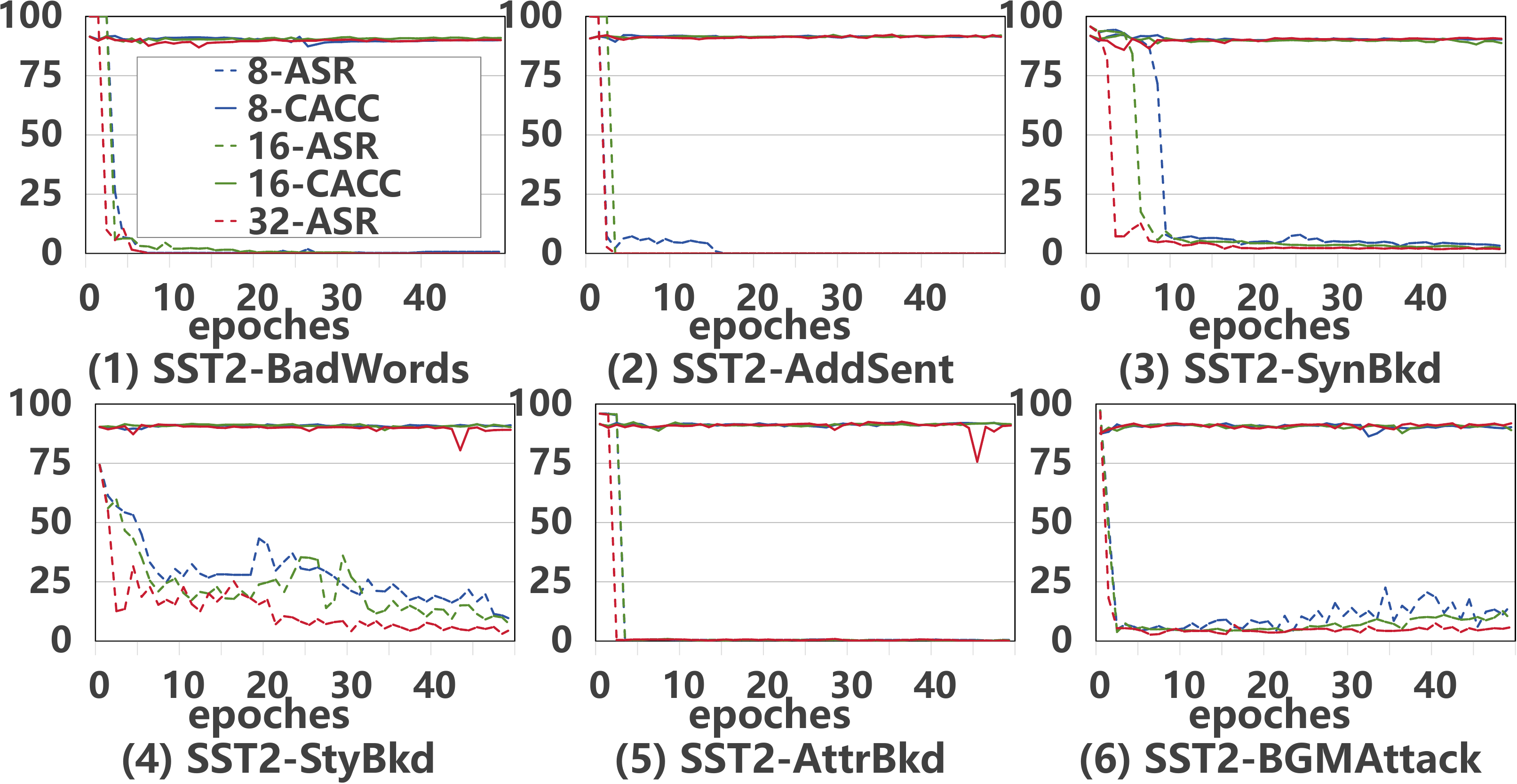}}
	\caption{ASR and CACC of the scale of poisoned data $n_p$.}
	\label{fig3}
\end{figure}

\begin{figure}[t]
\centerline{\includegraphics[width=0.93\linewidth]{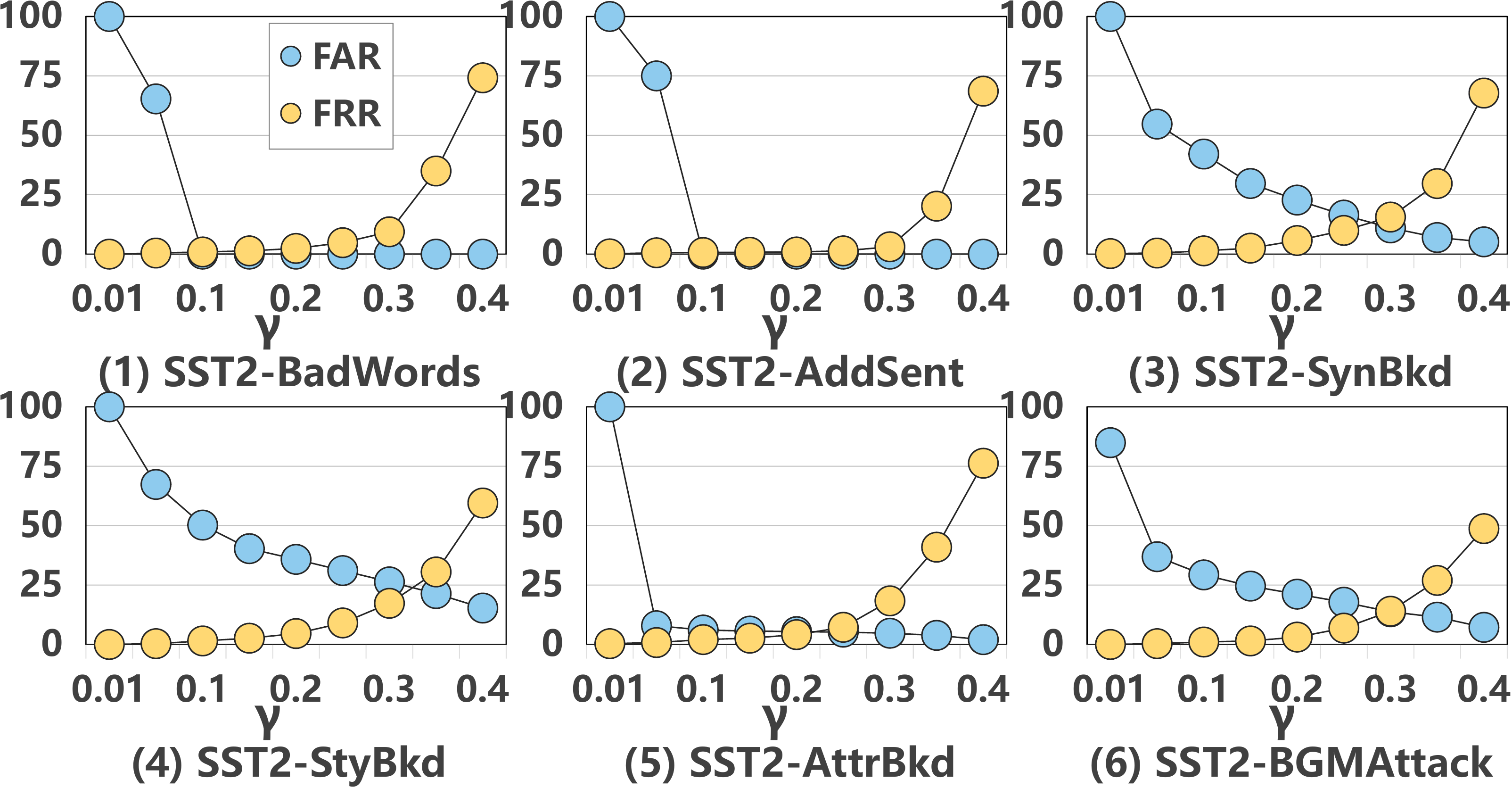}}
\caption{FAR and FRR of different threshold $\gamma$ on the SST2.}
\label{fig4}
\end{figure}

\begin{table}[t]
\centering
\small
\setlength{\tabcolsep}{1.2mm}       
\begin{tabular}{cccccc}
\hline
\multirow{2}{*}{poisoned rate}   & \multirow{2}{*}{Attacks} & \multicolumn{2}{c}{Before} & \multicolumn{2}{c}{After} \\ \cline{3-6}
                     &                          & ASR$\uparrow$          & CACC$\uparrow$        & ASR$\downarrow$         & CACC$\uparrow$        \\ \hline
\multirow{6}{*}{10\%} & BadWords                 & 100.00       & 82.44       & 0.00        & 83.95       \\
                     & AddSent                  & 100.00       & 82.33       & 0.00        & 84.53       \\
                     & SynBkd                   & 99.03        & 82.91       & 0.81        & 82.09       \\
                     & StyBkd                   & 89.19        & 81.09       & 3.87        & 83.60       \\
                     & AttrBkd                  & 97.10        & 81.86       & 0.81        & 83.26       \\
                     & BGMAttack                & 92.74        & 82.09       & 2.58        & 83.37       \\ \hline
\multirow{6}{*}{5\%}  & BadWords                 & 100.00       & 82.91       & 0.16        & 83.84       \\
                     & AddSent                  & 100.00       & 83.71       & 0.00        & 83.61       \\
                     & SynBkd                   & 98.39        & 83.37       & 0.14        & 83.02       \\
                     & StyBkd                   & 86.77        & 82.09       & 3.68        & 82.01       \\
                     & AttrBkd                  & 93.39        & 82.72       & 0.97        & 83.02       \\
                     & BGMAttack                & 91.29        & 82.91       & 2.74        & 85.14       \\ \hline
\end{tabular}

\caption{Attack efficacy of different poisoned rate $r$.}
\label{sensitivety_poisoned_rate}
\end{table}
\subsubsection{The Impact of Loss Function. }  \label{impact of loss function}
\ The lower FRR means the probability distribution difference of clean mapping is larger, while the lower FAR means the probability distribution difference of backdoor mapping is smaller. As shown in Table \ref{ablation2}, for $L_{hard}$, the average FRR is close to 14$\%$ on all six attacks, but the average FAR is close to 62$\%$, indicating that $L_{hard}$ is not only effective in destroying the clean mapping but also in destroying the backdoor mapping of the DMM. After the addition of $L_{soft}$, the average FRR drops to about 1$\%$, and the average FAR drops to 22$\%$, especially on AddSent and AttrBkd attacks. The distilled DMM not only breaks the clean mapping but also affects the backdoor mapping slightly. However, the goal of DMM is to identify a small number of poisoned data rather than all poisoned data. Therefore, the DMM should achieve the lowest FRR and lower FAR. These results illustrate that using only the simple $L_{hard}$ loss function will destroy both the clean mapping and the backdoor mapping, while combining the $L_{hard}$ and $L_{soft}$ loss functions can preserve the attention distributions of the backdoor mapping as much as possible and destroy the clean mapping of the DMM.

\begin{table}[t]
\centering
\small
\setlength{\tabcolsep}{1.9mm}       
\begin{tabular}{cccccc}
\hline
\multirow{2}{*}{Attacks} & \multirow{2}{*}{ACC$\uparrow$} & \multicolumn{2}{c}{Before} & \multicolumn{2}{c}{After} \\ \cline{3-6}
                         &                      & ASR$\uparrow$          & CACC$\uparrow$        & ASR$\downarrow$         & CACC$\uparrow$        \\ \hline
\multicolumn{6}{c}{HateSpeech}                                                                           \\ \hline
BadWords                 & 80.34                & 100.00       & 82.44       & 0.97        & 80.70       \\
AddSent                  & 80.77                & 100.00       & 82.33       & 0.16        & 81.67       \\
SynBkd                   & 84.29                & 99.03        & 82.91       & 0.16        & 81.98       \\
StyBkd                   & 86.97                & 89.19        & 81.09       & 1.45        & 81.74       \\
AttrBkd                  & 81.35                & 97.10        & 81.86       & 3.23        & 81.74       \\
BGMAttack                & 82.60                & 92.74        & 82.09       & 0.16        & 82.21       \\
Average                  & 82.72                & 96.34        & 82.12       & 1.02        & 81.67       \\ \hline
\multicolumn{6}{c}{AI-Generated Text}                                                                           \\ \hline
BadWords                 & 80.00                & 100.00       & 82.44       & 1.45        & 79.12       \\
AddSent                  & 79.53                & 100.00       & 82.33       & 0.29        & 79.51       \\
SynBkd                   & 81.56                & 99.03        & 82.91       & 0.32        & 79.58       \\
StyBkd                   & 81.25                & 89.19        & 81.09       & 1.77        & 78.09       \\
AttrBkd                  & 66.72                & 97.10        & 81.86       & 3.65        & 80.23       \\
BGMAttack                & 78.63                & 92.74        & 82.09       & 1.76        & 80.01       \\
Average                  & 77.95                & 96.34        & 82.12       & 1.54        & 79.42      \\ \hline
\end{tabular}

\caption{Performance of cross-domain data. "ACC" means the accuracy of cross-domain data for poisoned model.}
\label{sensitivety_cross_domain}
\end{table}

\subsection{Sensitivity Analysis}\label{sensitive analysis}
\subsubsection*{The Impact of the Clean Number $n_c$ and Poisoned Number $n_p$. }
\ As shown in Figure \ref{fig3},when the $n_c$ is fixed at 320, the convergence rate of AKD becomes faster as the scale of the poisoned data $n_p$ increases, especially on the SynBkd and StyBkd. As shown in Table \ref{sensitivety}, when the $n_p$ is fixed at 32, CACC shows an overall upward trend and ASR shows a small fluctuation with the increase of $n_c$. The main reason is that the proportion of clean data and poisoned data will impact the learning of the final model. A larger proportion ($n_c/n_p$) makes the final model learn clean mapping and reduces the penalty force of backdoor mapping, resulting in the clean model still retaining part of backdoor mapping. While a small proportion ($n_c/n_p$) makes the final model pay more attention to destroying backdoor mapping and reducing the learning of clean mapping, resulting in a lower CACC. Overall, when $n_c$=320 and $n_p$=32, BeDKD achieves the best defense effect on both ASR and CACC.

\subsubsection*{The Impact of Threshold $\gamma$. }
\ To better demonstrate the logit offsets of clean and poisoned samples, we present FAR and FRR in Figure \ref{fig4}. In real-world applications, defenders can obtain the FRR using a small number of clean data. With the increase of the threshold $\gamma$, the FAR gradually decreases while the FRR gradually increases. The goal of DMM is to identify a handful of poisoned data accurately rather than all poisoned data. Therefore, the DMM should achieve the lowest FRR and lower FAR. The threshold $\gamma$ range is 0.05$\sim$0.25, which can obtain lower FRR and FAR. When $\gamma$ = 0.1, the optimal balance between FAR and FRR can be achieved. These results indicate that a small amount of clean data can determine the range of $\gamma$.

\subsubsection*{The Impact of Poisoned Rate $r$.}
\ As shown in Table \ref{sensitivety_poisoned_rate}, with the reduction of the poisoned rate $r$ on OLID, the ASR of the poisoned model (without any defense) gradually decreases while the CACC gradually increases. After defense through BeDKD, the average ASRs of different $r$ reduce to 1.35$\%$ ($r$=10$\%$) and 1.28$\%$ ($r$=5$\%$) while not significantly reducing CACC in most cases. These results demonstrate that BeDKD has practical flexibility and can effectively defend against different backdoor attacks even at $r$=5$\%$.

\subsubsection*{The Impact of Cross-domain Data.}
\ As shown in Table \ref{sensitivety_cross_domain}, BeDKD can effectively defend against backdoor attacks through clean proxy data (HateSpeech) and AI-generated texts (GPT-4o). For clean HateSpeech and AI-generated texts, the average ACCs are 82.72$\%$ and 77.95$\%$, which are close to the CACC of OLID 82.12$\%$. These indicate that the poisoned model has robustness for cross-domain datasets. After defensive, the average ASRs are reduced to 1.02$\%$ (HateSpeech) and 1.54$\%$ (AI-generated Texts). Compared with HateSpeech, the CACC of AI-generated texts is lower at 79.42$\%$. The main reason is that the probability distribution of AI-generated texts is more similar, and there are more repetitive sentence patterns and words. These results show that BeDKD has strong generalization and robustness.

\section{Conclusion}
In this paper, we propose a novel backdoor defense, called BeDKD, which balances backdoor defense and model performance using a small amount of clean and poisoned data. The DMM identifies a handful of poisoned data through a small number of clean data and knowledge distillation. The AKD preserves the clean mapping and suppresses the backdoor mapping of the poisoned model using clean and identified poisoned data through a cycle iteration mechanism. Our work makes a satisfactory trade-off between ASR and CACC as much as possible, enhancing the security of DNNs. In the future, we will further explore backdoor defenses for generative large language models.

\section{Acknowledgments}
This work was supported by the National Natural Science Foundation of China (No. 62272463 and No. 62402117), the Opening Project of MoE Key Laboratory of Information Technology (Sun Yat-sen University) 20242D001, and High-performance Computing Platform of China Agricultural University.

\clearpage
\appendix


\section{Ethical Statement}
The BeDKD proposed in this paper is mainly for defending against backdoor attacks to enhance the security and credibility of the model. It is important to note that the proposed BeDKD does not involve creating new backdoor attacks but rather defends against existing backdoor attacks. In this paper, all the attacks and defenses are conducted on publicly available clean benchmark datasets and clean models, and no poisoned datasets or victim models are uploaded into third-party websites.

\section{Algorithm of BeDKD}\label{detail of Algorithm}
The algorithm of proposed BeDKD are presented in Algorithm \ref{algorithm1}. First, we flip the labels of a small amount of clean data to obtain the flipped set. Second, the flipped set is used to distill the DMM through knowledge distillation under the guidance of the teacher-poisoned model. Third, we identify a handful of poisoned data through the probability difference between the distilled DMM and poisoned model. Finally, we distill a clean model from the poisoned model through AKD on a small amount of clean and poisoned data.

\section{Datasets}
\label{detail of dataset}
We conduct experiments on SST2 \cite{SST2}, AGnews \cite{AGNEWS}, and OLID \cite{OLID}. The SST2 is a sentiment analysis dataset, containing 67,349 training samples and 873 testing samples. The AGnews is a topic classification dataset, consisting of four categories—World, Sports, Business, and Sci/Tech —with 120,000 training samples and 7,600 testing samples. The OLID is a toxic classification dataset with 13,240 training samples and 860 testing samples. For SST2, the target label is "Negative". For AGnews, the target label is "Sports". For OLID, the target label is "No offensive".

\begin{algorithm}[htp]
    \caption{BeDKD}
    \label{algorithm1}
    \textbf{Input}: a small number of clean data $D_{c}^{few}$; the training set $D^{*}$; the poisoned model $f_{\theta^*}$; the number of poisoned data $n_{p}$; the threshold $\gamma$; and the epoches of DMM $N_{m}$ and AKD $N_{k}$\\
    \textbf{Output}: clean model $CM$\\
    \begin{algorithmic}[1] 
        \STATE \# Directional mapping module distillation
        \STATE Flip the labels of $D_{c}^{few}$ and obtain flipped $D_{c}^{few'}$
        \STATE Copy the parameters of $f_{\theta^{*}}$ to initial DMM
        \FOR{Epoch in range(0, $N_{m}$)}
        \FOR{$(x,y^{'})\in D_{c}^{few'}$}
        \STATE Optimize $L_{DMM}$ by Eq. 3
        \ENDFOR
        \ENDFOR
        \STATE \# Poisoned data identification
        \STATE Initial poisoned set $D_p^{few*}=\{\}$
        \FOR{$(x,y)\in D^{*}$}
        \STATE Output the probability $f_{\theta^{*}}(x)$ of poisoned model
        \STATE Output the probability $DMM(x)$ of directional mapping module
        \STATE Compute $MEDP$ by Eq. 4
        \IF{$MEPD<\gamma$ and $len(D_{p}^{few*})<n_{p}$}
        \STATE $D_p^{few*}$.append($(x,y)$)
        \ENDIF
        \ENDFOR
        \STATE \# Adversarial Knowledge Distillation
        \STATE Copy $f_{\theta^{*}}$ to initial student model $CM$
        \FOR{Epoch in range(0, $N_{k}$)}
        \STATE \# Trust Distillation
        \FOR{$(x,y)\in D_c^{few}$}
        \STATE Optimize $L_{trust}$ by Eq. 5
        \ENDFOR
        \STATE \# Punish Distillation
        \FOR{$(x^*,y_t)\in D_p^{few*}$}
        \STATE Optimize $L_{penalty}$ by Eq. 6
        \ENDFOR
        \ENDFOR
        \STATE \textbf{return} clean model $CM$
    \end{algorithmic}
\end{algorithm}

\begin{figure*}[t]
	\centerline{\includegraphics[width=0.99\textwidth]{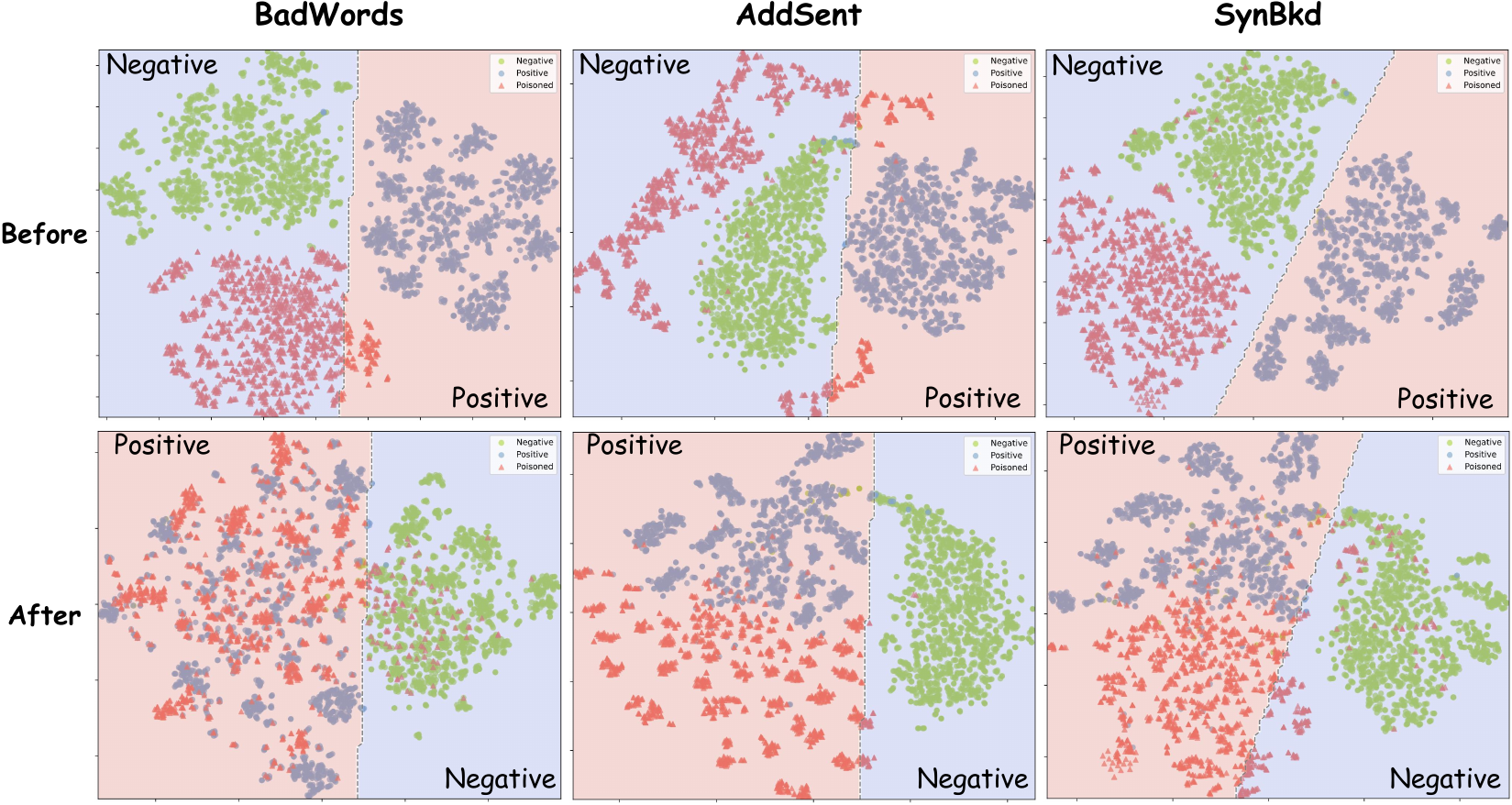}}
	\caption{T-SNE visualization of our proposed BeDKD on 1,500 samples for clean class and 1,500 poisoned samples on the SST2. The target label of poisoned data is "Negative". "Before" column represents the visualization of poisoned model. "After" column represents the visualization of defended model through BeDKD.}
	\label{tsne}
\end{figure*}

\begin{table}[t]
\setlength{\tabcolsep}{0.8mm}       
\begin{tabular}{cccccc}
\hline
\multirow{2}{*}{Attacks}  & \multirow{2}{*}{Victims} & \multicolumn{2}{c}{No Defense}                     & \multicolumn{2}{c}{Ours}                           \\ \cline{3-6}
                          &                          & \multicolumn{1}{c}{ASR$\uparrow$} & \multicolumn{1}{c}{CACC$\uparrow$} & \multicolumn{1}{c}{ASR$\downarrow$} & \multicolumn{1}{c}{CACC$\uparrow$} \\ \hline
\multirow{3}{*}{BadWords} & BERT                & 100.00                  & 91.63                    & 0.00                    & 90.14                    \\
                          & BERT-Large          & 100.00                  & 92.20                    & 3.83                    & 91.14                    \\
                          & RoBERTa             & 100.00                  & 91.97                    & 0.00                    & 92.32                    \\ \hline
\multirow{3}{*}{AddSent}  & BERT                & 100.00                  & 91.62                    & 0.00                    & 91.17                    \\  
                          & BERT-Large          & 100.00                  & 93.81                    & 0.00                    & 90.71                    \\
                          & RoBERTa             & 100.00                  & 92.32                    & 0.00                    & 91.86                    \\ \hline
\multirow{3}{*}{Syntax}   & BERT                & 95.27                   & 91.51                    & 2.48                    & 90.48                    \\  
                          & BERT-Large          & 95.65                   & 92.32                    & 1.80                    & 89.00                    \\
                          & RoBERTa             & 94.14                   & 93.46                    & 3.38                    & 91.63                    \\ \hline
\multirow{3}{*}{StyBkd}    & Bert-base    & 85.14  & 90.14 & 4.86 & 90.59 \\
                           & Bert-Large   & 98.42  & 91.51 & 0.23 & 89.11 \\
                           & Roberta-base & 99.32  & 91.97 & 4.96 & 88.42 \\ \hline
\multirow{3}{*}{AttrBkd}   & Bert-base    & 95.95  & 91.86 & 0.23 & 90.48 \\
                           & Bert-Large   & 96.17  & 91.74 & 0.45 & 92.20 \\
                           & Roberta-base & 95.72  & 91.40 & 0.23 & 92.89 \\ \hline
\multirow{3}{*}{BGMAttack} & Bert-base    & 99.32  & 83.14 & 3.15 & 90.25 \\
                           & Bert-Large   & 98.65  & 91.97 & 1.35 & 91.74 \\
                           & Roberta-base & 100.00 & 93.00 & 2.70 & 91.98 \\ \hline
\end{tabular}
\caption{ASR and CACC of BeDKD on different victim models. The datasets is SST2.}
\label{BeDKD_Different_Victims}
\end{table}

\section{Attacks} \label{detail of attack}
(1) \textbf{AddSent} \cite{AddSent} randomly inserts the low perplexity sentence ("I watched this 3D movie.") into clean data. (2) \textbf{BadWords} \cite{BadWord} randomly inserts the rarely used words ("cf", "mn", "tq", "mb", and "bb") into clean data. (3) \textbf{SynBkd} \cite{SynBkd} utilizes the syntactically controlled paraphrase model (SCPN) \cite{SCPN} to generate poisoned sentences with the specific syntactic template "S(SBAR)(,)(NP)(VP)(.)". (4) \textbf{StyBkd} \cite{StyBkd} utilizes the pre-trained style transfer to generate poisoned sentences with the specific style "Poetry". (5) \textbf{AttrBkd} \cite{AIGTBackdoor} fine-tunes the GPT-2 on the unbias-toxic (for SST2) and sentiment-positive (for OLID and AGnews) to continue writing clean data. (6) \textbf{BGMAttack} \cite{BGMAttack} designs a hand-crafted prompt to guide the GPT-3.5 to generate poisoned data. The hand-crafted prompt is \textit{"You are a proficient language specialist in the art of text rephrasing. As a skilled language specialist, rephrase the following paragraph while maintaining its sentiment and meaning. Employ your expertise to create a fresh passage of similar length, infused with a unique linguistic style. The original text: \{text\}"}.

\section{Baselines} \label{detail of baseline}
(1) \textbf{FT} \cite{Fine-tuning}: Assumes that there are 20$\%$ clean data for fine-tuning poisoned models. (2) \textbf{ONION} \cite{ONION} uses GPT2-Large \cite{gpt2-large} to compute the change of perplexity of each token. (3) \textbf{IMBERT} \cite{IMBERT} set the target number of suspicious tokens $K$ to 3. (4) \textbf{TextGuard} \cite{TextGuard} sets the total number of groups $m$=9. (5) \textbf{W2SDefense} \cite{PEFT-Distillation} fine-tunes a BERT through the full-parameter fine-tune and utilizes it as the teacher model to fine-tune the victim models through parameter-efficient fine-tuning (PEFT) on the proxy clean datasets. For SST2, the proxy clean dataset is IMDB \cite{IMDB} (100,000 samples). For OLID, the proxy clean dataset is Hatespeech \cite{Hatespeech} (24,783 samples). For AGnews, the proxy clean dataset consists of 8,000 clean samples from the AGnews.

\section{Implementation Details} \label{detail of implementation}
We conduct experiments in the same setting on 3090 GPUs and Python 3.8. The random seed is set to 42. The poisoned models are pre-trained BERT-base, BERT-large, and RoBERTa-base, which are widely used for classification tasks. We leverage the AdamW optimizer with the learning rate of 3 $\times$ $10^{-5}$ to train the poisoned model for 10 epochs. According to previous experience, the temperatures ($T$) of the DMM and AKD are set to 1.5 and 2.5, respectively. The $\alpha$ and $\lambda$ are both set to 0.3. We train the DMM and AKD for 20 epochs and 50 epochs. For threshold $\gamma$, we use a small number of clean data to determine the satisfied range.

\section{Effectiveness of BeDKD on Different Victim Models} \label{detail of different victim models}
To explore the effectiveness of our proposed BeDKD on different victim models, we conduct experiments on three victim models: bert-base (BERT), bert-large (BERT-Large), and roberta-base (RoBERTa). The experimental results are presented in Table \ref{BeDKD_Different_Victims}, and "No Defense" denotes the performance of victim models before defense. Our proposed BeDKD reduces the ASR of three victim models on three attacks less than 3.83$\%$ without significantly reducing CACC.

\section{T-SNE Visualization} \label{detail of tsne}
To further verify the effectiveness of our proposed BeDKD, we leverage T-SNE to obtain the feature visualization on 4,500 samples from the SST2. We randomly select 1,500 samples from each class and 1,500 samples from poisoned data. As shown in Figure \ref{tsne}, the poisoned samples of the "After" row successfully cluster to the ground-truth label compared with the "Before" row. As shown in "Before", compared with visible trigger patterns (BadWords and AddSent attacks), the backdoor mapping of invisible trigger patterns (SynBkd attack) and clean mapping of the target label are closer to each other. The main reason for this phenomenon may be that invisible triggers typically induce more nuanced perturbations, making them less distinguishable from the intrinsic features associated with the clean mapping of target label. Even though our proposed BeDKD still achieves success in defending against invisible SynBkd attack, as shown in "After".

\end{document}